\documentclass[aps,reprint,12pt]{revtex4-1}
\usepackage[version=3]{mhchem} 
\usepackage[hidelinks]{hyperref}
\usepackage{graphicx}
\usepackage{amsfonts,amsmath,amssymb,bm}
\usepackage{color}
\usepackage{multirow}
\usepackage[utf8]{inputenc}
\usepackage{wrapfig}
\usepackage{xfrac}
\usepackage{bm}
\graphicspath{{./figures/}}
\begin{document}

\title{Comparing Meta-GGAs, +U Corrections, and Hybrid Functionals for Polaronic Point Defects in Layered MnO$_2$, NiO$_2$, and KCoO$_2$}


\author{Raj K. Sah}
\affiliation{Department of Physics, Temple University, Philadelphia, PA 19122}

\author{Michael J. Zdilla}
\affiliation{Department of Chemistry, Temple University, Philadelphia, PA 19122}

\author{Eric Borguet}
\affiliation{Department of Chemistry, Temple University, Philadelphia, PA 19122}

\author{John P. Perdew}
\affiliation{Department of Physics and Engineering Physics, Tulane University, New Orleans, LA 70118}

\begin{abstract}
Defects in a material can significantly tune properties and enhance utility. Hybrid functionals like HSE06 are often used to describe solids with such defects. However, geometry optimization (including accounting for effects such as Jahn-Teller distortion) using hybrid functionals is challenging for the large supercells needed for defect study. The proposed r$^2$SCAN+rVV10+U+$\mathrm{U_d}$ method, which is computationally much cheaper and faster than hybrid functionals, can successfully describe defects in materials with the proper choice of U (for the d orbitals of the host atom) and $\mathrm{U_d}$ (for those of the defect atom), as shown here for small polaron defects in layered transition-metal oxides. For a range of U and $\mathrm{U_d}$ around literature values (from solid-state reaction energies) for a given transition-metal ion and its oxidation state, we find that this approach predicts localized polaronic states in band gaps, as hybrid functionals do. The layered materials birnessite ($\mathrm{K_nMnO_2}, n=0.03 $) and $\mathrm{K_nNiO_2},n=0.03$, with one K atom intercalated between layers in a supercell, are found to have one localized occupied $\mathrm{e_g}$ polaronic state on the transition metal ion reduced by the insertion of the K atom, when the geometry is calculated as above using published U values. The expected Jahn-Teller distortion is not observed when U=$\mathrm{U_d}$=0. Layered cobalt oxide with additional potassium ions intercalated ($\mathrm{K_nCoO_2},n=1.03$) is  different, due to a dramatic difference in electronic configuration of the defected Co(II) ion: A single extra K atom in the supercell leads to four localized electrons in the band gap, using standard U values, and  even for U=$\mathrm{U_d}$=0.
\end{abstract}
\maketitle{}
\noindent
\section{Introduction}
This work is part of a larger project on the computational identification of alkali-atom-intercalated  layered materials as promising catalysts for the oxygen evolution reaction (OER) of water splitting for clean hydrogen production. Earlier work \cite{peng-2017,Ding-2021} showed the importance of small-polaron defects formed by transfer of an electron from an intercalated alkali atom to a neighboring manganese ion. Electron flow in catalysis seems to be facilitated by a localized polaronic state in the energy gap close to the bottom of the conduction band. Calculation of the polaron often requires a nonlocal density functional for the exchange-correlation energy, and geometry optimization in a large supercell. We found that hybrid functionals as used in Ref. \cite{peng-2017} are too expensive for a broad materials search. In this article, we show that an alternative r$^2$SCAN+rVV10+U+$\mathrm{U_d}$ approach can predict localized polaronic states at much lower cost, and may be useful for other point defects in a range of materials. We also found that, while one intercalated K atom creates one localized electron in the energy gap of layered MnO$_2$ and NiO$_2$, it can create four in layered KCoO$_2$. We further found that, while r$^2$SCAN needs a +U correction to create a polaron in MnO$_2$ and NiO$_2$, it does not need one in KCoO$_2$. Our broader computational and experimental materials search is now underway.

Various types of defects exist in solids, and defects in solids can influence numerous important properties like electrical conductivity, reactivity, and magnetic or optical properties. For example, the Oxygen Evolution Reaction (OER) is frequently favored by defects like polarons in transition metal oxide (TMO)-based catalysts \cite{peng-2017,Ding-2021}. Leveraging defects as a tool allows fine tuning of the electronic properties of materials, making the understanding of defects in materials a pivotal area of research. 

The computational design and study of such materials using first-principles density functional theory (DFT) \cite{HK-1964,KS-1965} offer valuable early insights. However, the approach presents challenges, as density functional approximations (DFAs) introduce self-interaction error (SIE). Popular DFAs such as LDA/GGA/meta-GGA tend to underestimate the Perdew-Parr-Levy-Balduz (PPLB) \cite{PPLB-1961} straight line condition, leading to inaccuracies in describing charge transfer that are reduced but not eliminated in that sequence of functionals. Given that defect studies require proper charge transfer, DFAs often fall short in accurately portraying defects in a system.

\begin{figure*}
    \centering
    \resizebox{0.7\textwidth}{!}{\includegraphics{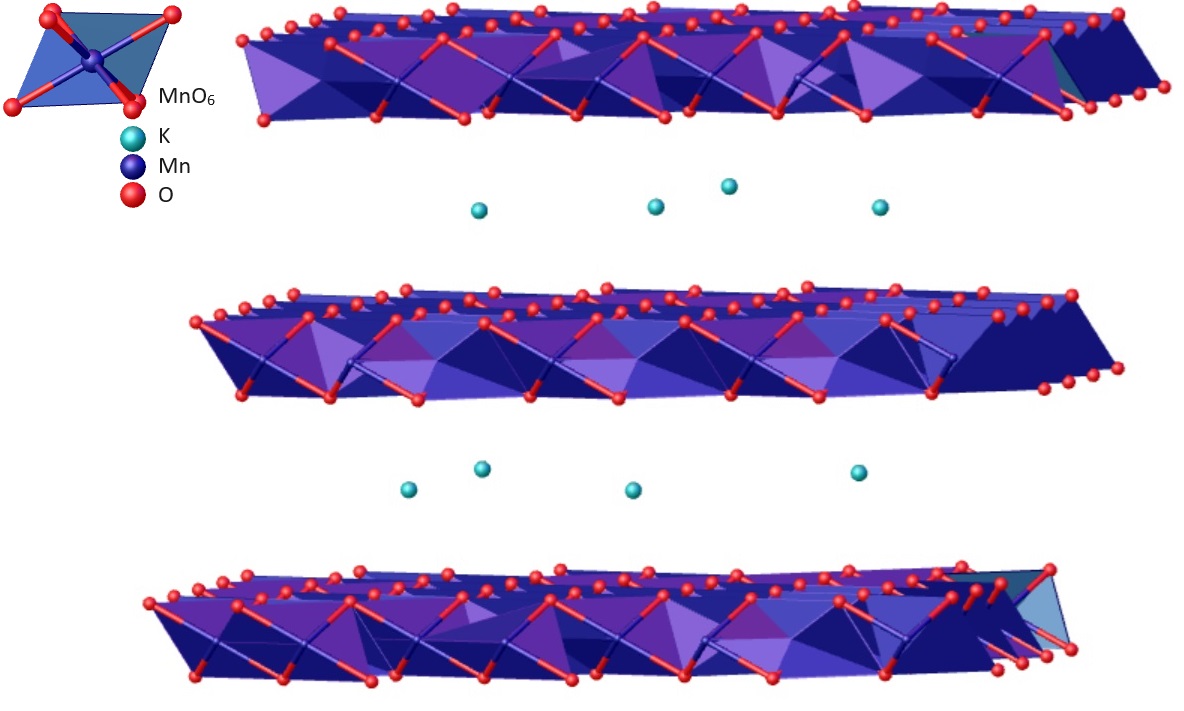}}
    \caption{General structure for layered potassium metal oxides $\mathrm{K_nMnO_2}$.}
    \label{fig:figurex1}
\end{figure*}

Hybrid functionals, which combine a fraction of exact exchange, such as HSE06 \cite{HSE06-Krukau-2006} which utilizes 25\%  exact exchange and 75\% PBE exchange in the short range along with full PBE exchange in the long range,  experience less SIE and often provide a more accurate description of the electronic structure of materials. Hybrid functionals have been successfully employed in studying defects in solids. For instance, Peng et al. \cite{peng-2017} effectively investigated polaron-like defects in birnessite (Fig. \ref{fig:figurex1}, $\mathrm{K_nMnO_2},n<1$). Nevertheless, the inclusion of exact exchange in hybrid functionals renders them computationally expensive. The structural relaxation using hybrid functionals becomes particularly costly for a reasonably sized supercell with localized defects, with computational expenses rapidly escalating as the supercell size increases toward better simulation of defects. Consequently, the structural optimization of large supercells becomes impractical. Another challenge associated with hybrid functionals is that a material-independent exact exchange mixing parameter is not determined through any exact condition, nor is the range-separation parameter in range-separated hybrids \cite{sun-2016}. Peng et al. determined a mixing parameter of 0.22 to study defects in birnessite using HSE06 \cite{peng-2017}, deviating from the original value of 0.25. Additionally, determining a mixing parameter for semiconductors that may not be suitable to metals or insulators \cite{scuseria-2011} adds to the challenge of finding an appropriate parameter for the system under study when employing hybrid functionals.

$\mathrm{r^2SCAN}$ \cite{sun-2020} is a recently developed meta-generalized gradient approximation that reinstates exact constraint adherence to rSCAN \cite{yates-2019}, preserving the numerical efficiency of  rSCAN while simultaneously restoring the  transferable accuracy of  SCAN \cite{SCAN}. Several studies have demonstrated that SCAN predicts geometries and other properties as well as or even better than hybrid functionals. Sun et al. \cite{sun-2016} showed that SCAN accurately predicts geometries and energies of diversely bonded materials and molecules, matching or surpassing the accuracy of computationally expensive hybrid functionals. Another study by Sa\"{y}nick and Cocchi \cite{cocchi-2021} on cesium-based photocathode materials $\mathrm{Cs_3Sb}$ and $\mathrm{Cs_2Te}$ reported excellent performance of SCAN and HSE06 for both structural and electronic properties. A recent paper on the arXiv  \cite{wickramaratne-2023} reports that, while SCAN may not reliably describe the properties of deep defects and small polarons in several semiconductors and insulators, it yields remarkably good agreement with experimental structural parameters for materials like ZnO, GaN, $\mathrm{Ga_2O_3}$ and NaF. Additionally, a study by Varadwaj and Miyake \cite{varadwaj-2022} on the geometrical, electronic, and optical properties of vanadium dioxide found that SCAN and SCAN-rVV10 can adequately predict the most important geometrical and optoelectronic properties of $\mathrm{VO_2}$. Numerous related studies further support the idea that SCAN successfully describes the structure and other properties of materials. Given that $\mathrm{r^2SCAN}$ closely agrees with SCAN in accuracy, we expect that $\mathrm{r^2SCAN}$ would exhibit similar accuracy in the aforementioned studies.

By construction, $\mathrm{r^2SCAN}$ can exhibit considerably less self-interaction error (as reflected in its smaller U value \cite{Gautam2018}) than standard GGAs. (The r$^2$SCAN effective one-electron potential can still show substantial 
self-interaction error.) In this study, we demonstrate that the $\mathrm{r^2SCAN+rVV10+U+U_d}$ functional can predict the existence of localized polaronic states at defects, as the HSE06 hybrid functional does. Here, U represents the Hubbard U correction of Anisimov and collaborators \cite{anisimov-1991,anisimov-1993,anisimov-1994}, applied to transition metal sites other than the defect site, while $\mathrm{U_d}$ is the correction applied to the defect site. Cococcioni and de Gironcoli \cite{Cococcioni-2005} showed that the +U correction can be regarded as a many-electron self-interaction correction that, like the PPLB condition \cite{PPLB-1961}, penalizes non-integer electron number on a localized orbital to which it is applied.

Additionally, vdW denotes the long-range van der Waal’s correction. In our approach, we utilize rVV10 \cite{Vydrov-2010,Sabatini-2013} to account for this interaction, which importantly reduces inter-layer spacing. In many of our calculations, including HSE06 without vdW, we use the supercell volume and shape from $\mathrm{r^2SCAN+rVV10+U+U_d}$, although we can as a check relax the internal coordinates with HSE06. 

Transition metal ions that serve as sites for defect formation in solids, are in different formal oxidation states (OS) than otherwise-identical  ions. For example, in a K-intercalated $\mathrm{MnO_2}$, the polaronic Mn site is in the +3 OS and the remaining Mn ions are in the +4 oxidation state \cite{peng-2017,Ding-2021}. Ions in different OS have different numbers of d electrons. Ref. \cite{Long-2020} reports that the ideal U correction decreases with increasing OS, which is attributed to a lower number of exchange interactions among fewer d electrons in a higher oxidation state. This behavior has been observed and reported \cite{Long-2020} for vanadium ions, and suggests U$\mathrm{_d} >$ U.

However, this situation is not universal, as we can see that U values for Mn for $\mathrm{r^2SCAN}$ are 1.5 eV, 2.1 eV and 1.8 eV for Mn ions with average oxidation states +3.5, +2.5 and +2.33 respectively \cite{Swathilakshmi-2023}. Here we see that Mn in the +2.5 average OS needs more U correction than Mn in the +2.33 average OS, which does not follow the trend observed for vanadium, although U for Mn in the +3.5 average OS supports the trend within the same elemental series. However it is obvious for the Mn oxide system that Mn ions in different oxidation states need different U corrections. 
Another example for Co is found in Ref. \cite{Long-2020}, which found U=3.0 eV as a correction to SCAN for  the oxidation reaction 6CoO+$\mathrm{O_2 \rightarrow 2Co_3O_4}$ (in which the Co ions are in the +2 and +2.67 oxidation states), but also found that SCAN with this U wrongly predicted the meta-stable crystal $\mathrm{O1-CoO_2}$ (in which the Co oxidation state is +4) to be non-metallic, while SCAN with U=0 correctly predicted it to be a metal. Thus the U value can vary with the oxidation state of ions in the system. This indicates that if a system has transition metal ions in different OS, the proper way to describe such a system would be to apply different U corrections to the transition metal ions based on their OS. This could be because DFAs make different SIE for ions in different OS.

Here we present a study of defects in solids by the double U correction $\mathrm{r^2SCAN+rVV10+U+U_d}$ method, where $\mathrm{U_d}$ is the U correction applied to the defect site and U is the U correction applied to remaining sites as required for the system. Since U typically increases with decreasing OS, and defect sites are typically in a lower oxidation state, we expect $\mathrm{U_d \geq U}$, although exceptions are possible.

One can use the $\mathrm{r^2SCAN+rVV10+U+U_d}$  geometry and the HSE06 hybrid functional orbital energies to leverage the best features of the hybrid functional (e.g., band gap). It appears that employing the full $\mathrm{r^2SCAN+rVV10+U+U_d}$  and/or HSE06 using $\mathrm{r^2SCAN+rVV10+U+U_d}$   geometry  is a state-of-the-art method for efficient study of defects in solids.

From a chemical perspective \cite{PAtkins}, for electronic configurations (3d)$^n$ $(4\leq n\leq 7)$, the spin state of a transition metal ion M in an MO$_6$  motif arises from a competition between crystal field effects which favor the low-spin state and internal effects which favor the high-spin state. The crystal field effect (which in the absence of spin removes the degeneracy of three t$_{2g}$ and two e$_g$ orbitals) depends on the lattice geometry, including any Jahn-Teller distortion of the MO$_6$ octahedron. The t$_{2g}$ orbitals have lobes between the bond axes, minimizing the repulsion between the metal ion and its surrounding oxygen ions. The ions we consider here are in oxidation states Mn(IV) (d$^3$), Mn(III) (d$^4$), Ni(IV) (d$^6$), Ni(III) (d$^7$), Co(III) (d$^6$), and Co(II) (d$^7$).

\section{Results and Discussion}
We choose three layered pristine TMOs, $\mathrm{MnO_2}$, $\mathrm{NiO_2}$, and $\mathrm{KCoO_2}$ as the starting point for our study.  Transition metal and oxygen ions in these TMOs are arranged in layers of MO6 (where M=Ni, Mn, Co) octahedra, with 32 transition metal ions per supercell. Inserting an additional potassium ion into the supercell between the layers creates a polaronic defect, specifically a Jahn-Teller electron small polaron \cite{peng-2017}. This defect has been studied in birnessite ($\mathrm{MnO_2}$) using the HSE06 functional \cite{peng-2017}. 

\subsection{Layered MnO$_2$}
\begin{figure*}
    \centering
    \resizebox{0.7\textwidth}{!}{\includegraphics{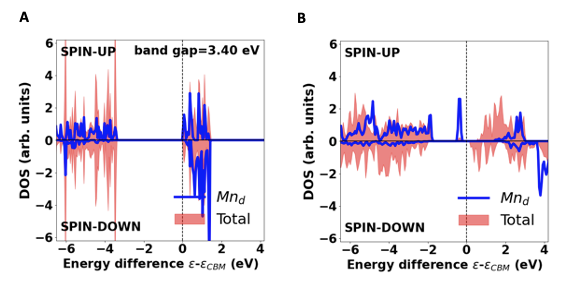}}
    \caption{HSE06 ($\alpha$ = 0.22) spin-resolved density of states per atom in (A) pristine MnO$_2$ with Mn(IV) ions and (B) MnO$_2$ with a single intercalated K$^+$ ion and a defect Mn(III) cation in a supercell with 97 atoms. For comparison with Ref. \cite{peng-2017}, the interlayer spacing is set to 7.12 Å, and internal coordinates are relaxed in HSE06. In Figs. 1-9 (except in Figs. \ref{fig:figure4}, and \ref{fig:figure7}), the shaded area shows the total density of one-electron states, and the blue curve shows the transition-metal-d states projected onto the site where the small polaron forms or is expected to form. Also in Figs. 1-9 (except in Figs. \ref{fig:figure4}, and \ref{fig:figure7}), the peaks (if any) in the band gap below the conduction band minimum (CBM) are occupied, localized states of the defect ion. The energy zero is set to the energy of the lowest-energy unoccupied orbital. The total density has been scaled down to make it comparable in size to the projected Mn-d density of states.}
    \label{fig:figure1}
\end{figure*}

First, we calculated the one-electron density of states (DOS) for both the pristine(i.e., non-intercalated) and a K-intercalated birnessite (Fig. \ref{fig:figurex1}) using the HSE06 functional with an exact exchange mixing parameter ($\alpha$) of 0.22 to reproduce  the work of  Peng et al. \cite{peng-2017}. The resulting DOS are shown in Fig. \ref{fig:figure1}. Comparing Figs. \ref{fig:figure1}A and \ref{fig:figure1}B in the current work with Figs. 2B and 2C of Peng et al \cite{peng-2017}, we conclude that we have successfully reproduced the latter’s DOS results. In Fig. \ref{fig:figure1}B, a K-intercalated birnessite exhibits the appearance of Mn(III) d-states at higher energy, a polaronic peak at the conduction band (CB) edge, and a break in  spin symmetry of the total DOS in the CB. These effects  were previously seen and explained by Peng et al. \cite{peng-2017}. The spin symmetry of the total DOS in pristine MnO$_2$ arises from the antiferromagnetic order of the Mn(IV) ions, and is disrupted by the defect Mn(III) ion (while the other pristine materials are non-magnetic). The appearance of the polaron peak is attributed to e$_g^1$ states in Mn(III) \cite{peng-2017,Ding-2021}.

\begin{figure*}
    \centering
    \resizebox{1.0\textwidth}{!}{\includegraphics{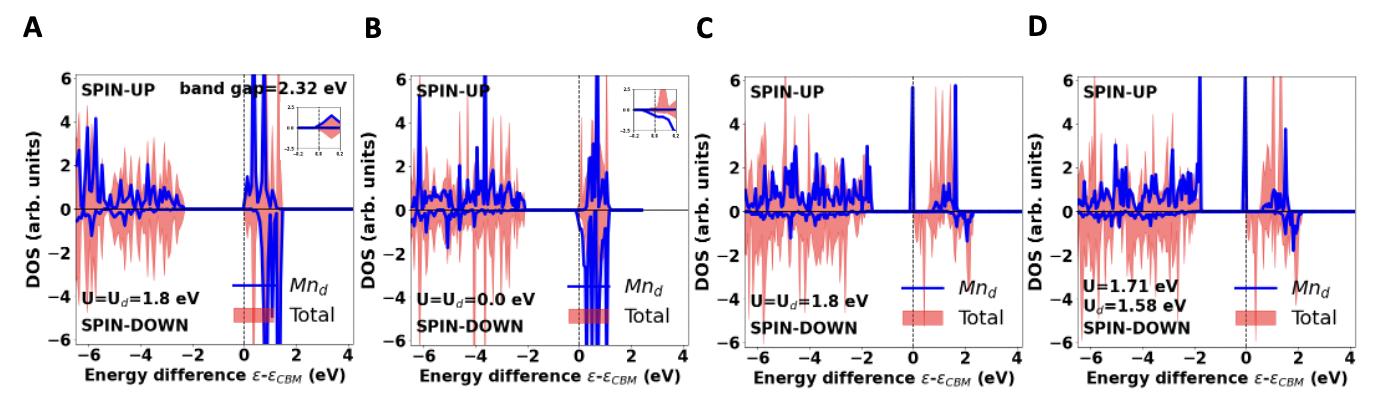}}
    \caption{r$^2$SCAN+rVV10+U+$\mathrm{U_d}$ spin-resolved density of states per atom in (A) pristine MnO$_2$ with U=$\mathrm{U_d}$=1.8 eV and in a single K-intercalated MnO$_2$ with (B) U = $\mathrm{U_d}$ = 0.0 eV, (C) U= $\mathrm{U_d}$ =1.8 eV, (D) U=1.71 eV and $\mathrm{U_d}$ =1.58 eV. The geometry has been optimized in r$^2$SCAN+rVV10+U+$\mathrm{U_d}$, which makes the interlayer spacing $\approx$} 5.26 Å. The insets in Figs. A and B show the DOS near the 0 eV region. The inset of Fig. B shows that for U=$\mathrm{U_d}$=0 the extra electron from the intercalated K goes to the bottom of the conduction band, making a semi-metal.
    \label{fig:figure2}
\end{figure*}

We will see similar effects upon a K-intercalation in $\mathrm{NiO_2}$ and $\mathrm{KCoO_2}$ systems later, and these effects can be understood through similar reasoning. The d-state electronic configurations of Mn(IV) in pristine $\mathrm{MnO_2}$ and Mn(III) in a single-K-intercalated $\mathrm{MnO_2}$ are as shown in Fig. \ref{fig:figure4} A and B respectively. 

\begin{figure*}
    \centering
    \resizebox{1.0\textwidth}{!}{\includegraphics{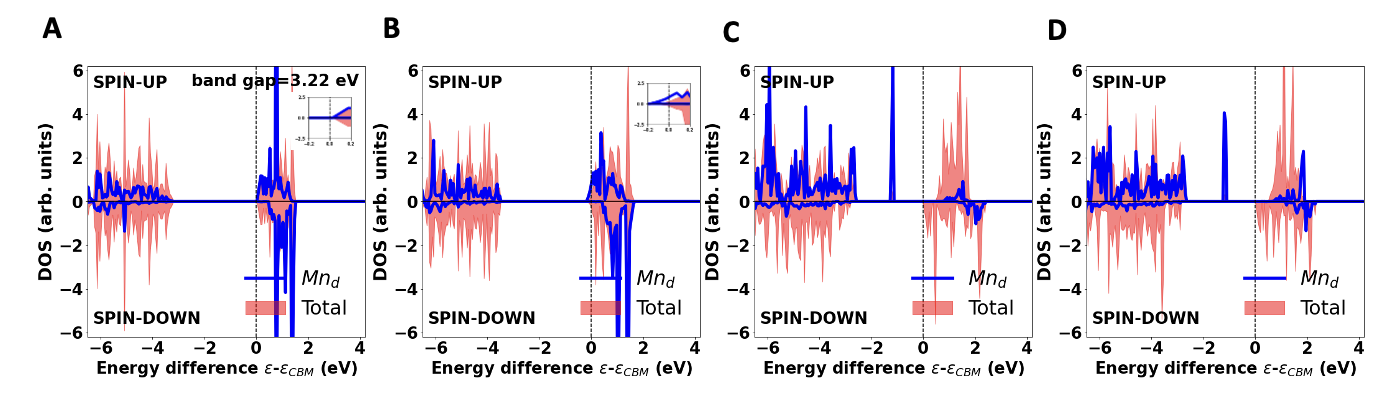}}
    \caption{HSE06($\alpha$=0.25)+D3  spin-resolved density of states per atom using r$^2$SCAN+rVV10+U+$\mathrm{U_d}$ geometry of Fig. \ref{fig:figure2} in (A) pristine MnO$_2$ with U=$\mathrm{U_d}$=1.8 eV and in a single K-intercalated MnO$_2$ with  (B) U=$\mathrm{U_d}$=0.0 eV, (C) U=$\mathrm{U_d}$=1.8 eV,  (D) U=1.71 eV and $\mathrm{U_d}$=1.58 eV. The insets in Figs. A and B show DOS near the 0 eV region. The inset in Fig. B shows that for U=$\mathrm{U_d}$=0 the extra electron from the intercalated K goes to the bottom of the conduction band, making a semi-metal.}
    \label{fig:figure3}
\end{figure*}
\begin{figure*}
  \centering
   \resizebox{0.8\textwidth}{!}{\includegraphics{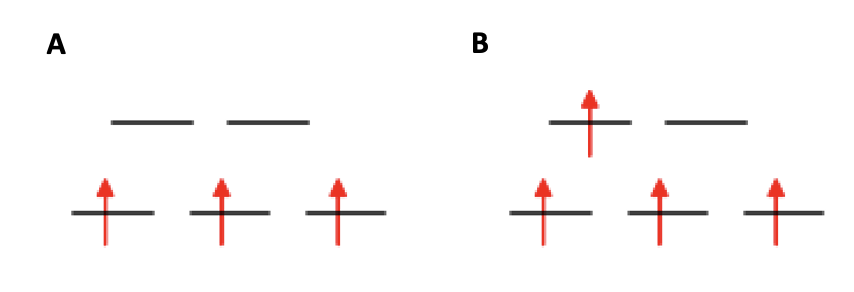}}
  \caption{d-orbital splitting (A) in Mn(IV) in pristine MnO$_2$ demonstrating  filled t$_{2g}$ states and empty e$_g$ states and (B) in Mn(III) in a single K-intercalated MnO$_2$.}
  \label{fig:figure4}
\end{figure*}
Swathilakshmi et al. \cite{Swathilakshmi-2023} recently determined the optimal U value for r$^2$SCAN for Mn to be 1.8 eV. They utilized three  oxidation  reactions:  MnO$\rightarrow$Mn$_2$O$_3$, MnO$\rightarrow$Mn$_3$O$_4$,  and Mn$_2$ O$_3$$\rightarrow$MnO$_2$. The U values for these reactions are 2.1 eV, 1.8 eV and 1.5 eV respectively \cite{Swathilakshmi-2023}. The average OS of Mn ions in these reactions are +2.5, +2.3 ,and +3.5 respectively \cite{Swathilakshmi-2023}. U values for these reactions are small compared to those for the PBE GGA and do not differ significantly, suggesting that r$^2$SCAN makes a small, comparable SIE but not an equal one for Mn(II), Mn(III) and Mn(IV) ions in manganese oxide systems.

First, we performed $\mathrm{r^2SCAN+rVV10+U}$ calculations for pristine birnessite with the optimal U of 1.8 eV and obtained the DOS as shown in Fig. \ref{fig:figure2}A. $\mathrm{r^2SCAN+rVV10+U}$ predicts a band gap of 2.30 eV for the pristine structure. The underestimation of the band gap compared to HSE06 by meta-GGA/GGA is a well-known general trend \cite{Brothers2008}. Next, we performed $\mathrm{r^2SCAN+rVV10+U+U_d}$ calculations for a K-intercalated MnO$_2$. As  $\mathrm{r^2SCAN}$ exhibits a small SIE,  we initially set U=$\mathrm{U_d}$=0.0 eV to see the performance of r$^2$SCAN+rVV10 without U correction. However, this method failed to resolve the Jahn-Teller defect, as depicted in Fig. \ref{fig:figure2}B. The corresponding electron from the K atom is delocalized, which can be seen by the extra electrons between the CB minimum and the chemical potential or Fermi level at 0 eV. This can be clearly seen in the inset of Fig. \ref{fig:figure2}B.  For the description of a polaron in a single K-intercalated MnO$_2$, DFA has to transfer charge from the inserted K atom to a Mn atom. The charge delocalization error of DFAs prohibits transferring a complete electron from the K atom to the Mn site.  This indicates the need for +U correction, which can remove the partial occupancy and localize the electron on the defect site. Subsequently, we tested the optimal U by setting U=$\mathrm{U_d}$=1.8 eV and the obtained DOS is shown in Fig. \ref{fig:figure2}C, revealing the appearance of a polaronic peak, and the appropriate distortion of Mn-O bonds at the Jahn-Teller distorted manganese atom (see Fig. S2 B). The +U correction applies an energy penalty to the partially occupied orbital.  As a result the polaron is localized and appears at the CB edge. We see shifting of Mn(III) d-states to higher energy, and spin symmetry breaking in the total DOS in the CB, consistent with the HSE06 results. As discussed in the work by Peng et al. \cite{peng-2017} and Ding et al. \cite{Ding-2021}, polaron formation can with the right distribution of intercalated potassium atoms create a potential step between layers that facilitates electron transfer between layers and enhances 
catalytic activity.

Motivated by studies indicating that U depends on OS, and that generally U decreases with an increase in OS, we determined U for the r$^2$SCAN+rVV10+U functional for Mn(III) and Mn(IV) by matching r$^2$SCAN+rVV10+U’s magnetic moment for a pristine material to the HSE06+D3 magnetic moment (Supplementary Material (SM) \cite{SM-defect}).  (All magnetic moments reported in this article arise from spin density inside the Wigner-Seitz sphere.) This method gave U values of 1.71 eV and 1.58 eV for Mn(III) and Mn(IV) ions, respectively. These values are close to the values in Ref. \cite{Swathilakshmi-2023}.

We then obtained the r$^2$SCAN+rVV10+U+$\mathrm{U_d}$ DOS of a K intercalated MnO$_2$ with U=1.71 eV and $\mathrm{U_d}$=1.58 eV which is as shown in Fig. \ref{fig:figure2}C. U=1.71 eV and $\mathrm{U_d}$=1.58 eV gave similar results to U=$\mathrm{U_d}$=1.8 eV. 

We also performed HSE06+D3 calculations on the r$^2$SCAN+rVV10+U+$\mathrm{U_d}$ geometries, and the resulting DOS are plotted in Fig. \ref{fig:figure3}.  The D3 dispersion correction was used for the HSE06 functional with new D3 damping parameters a1 = 0.383, a2 = 5.685, and s8 = 2.310 generated for the HSE06 functional \cite{Grimme-2014}. For HSE06+D3 calculations, we used an exact exchange mixing parameter  ($\alpha$) of 0.25. This is because we found that the r$^2$SCAN+rVV10+U geometry in much better agreement with the HSE06+D3 geometry  for $\alpha$=0.25 than for 0.22. More details are in the SM \cite{SM-defect}. 

Fig. \ref{fig:figure3}A shows the HSE06+D3 DOS of pristine MnO$_2$ obtained using the r$^2$SCAN+rVV10+U+$\mathrm{U_d}$ (U=$\mathrm{U_d}$=1.8 eV) geometry. HSE06+D3 increases the band gap of pristine MnO$_2$, close to the HSE06 result (Fig. \ref{fig:figure1}A). Fig. \ref{fig:figure3}B shows the HSE06+D3 DOS of a K-intercalated MnO$_2$ obtained using the r$^2$SCAN+rVV10+U+$\mathrm{U_d}$ (U=$\mathrm{U_d}$=0.0 eV) geometry. Like the r$^2$SCAN+rVV10+U+$\mathrm{U_d}$ (U=$\mathrm{U_d}$=0.0 eV) method, HSE06+D3 also fails to localize the defect state due to the absence of the localized geometric distortion near a single Mn atom. Fig. \ref{fig:figure3}C shows the HSE06+D3 DOS of a K-intercalated MnO$_2$ obtained using the r$^2$SCAN+rVV10+U+$\mathrm{U_d}$ (U=$\mathrm{U_d}$=1.8 eV) geometry. Here, the defect state appears as in the r$^2$SCAN+rVV10+U+$\mathrm{U_d}$ (U=$\mathrm{U_d}$=1.8 eV) method, but deep in the band gap. Fig. \ref{fig:figure3}D shows the HSE06+D3 DOS of a K-intercalated MnO$_2$ obtained using the r$^2$SCAN+rVV10+U+$\mathrm{U_d}$ (U=1.71 eV and $\mathrm{U_d}$= 1.58eV) geometry. Here, also, the defect state appears as in the r$^2$SCAN+rVV10+U+$\mathrm{U_d}$ (U=$\mathrm{U_d}$=1.8 eV) method, but deep in the band gap.

We also calculated HSE06 and HSE06+D3 DOS using an exact exchange mixing parameter $\alpha$ of  0.22 as determined by Peng et al. \cite{peng-2017} on r$^2$SCAN+rVV10+U+$\mathrm{U_d}$ geometry with U=$\mathrm{U_d}$=1.8 eV and obtained DOS as shown in Fig. S1 A and B, respectively. As expected, HSE06($\alpha$=0.22)  and HSE06($\alpha$=0.22)+D3 DOS do not look different. Notably, the HSE06($\alpha$=0.22) DOS does not look different from the full HSE06($\alpha$=0.25)+D3 evaluated on the same geometry, which is shown in Fig. \ref{fig:figure3}C.

R. Ding et al. \cite{Ding-2021} have proposed a position of the polaron close to the CB in  layered MnO$_2$ with an alternation of polaron-rich and polaron-poor layers. This scenario matches better with r$^2$SCAN+rVV10+U+$\mathrm{U_d}$ DOS, where the polaron is close to the CB. 

It is challenging to conclude which method provides a better description of the defect. However, we can assert that all three methods- HSE06, r$^2$SCAN+rVV10+U+$\mathrm{U_d}$ and HSE06+D3 using r$^2$SCAN+rVV10+U+$\mathrm{U_d}$ geometry- have successfully described the polaronic defect in birnessite with the proper U and $\mathrm{U_d}$ for r$^2$SCAN+rVV10+U+$\mathrm{U_d}$ calculations.

\subsection{Layered NiO$_2$}
Our next system under study was a layered NiO$_2$. This material has a hexagonal crystal structure with space group P6$_3$/mmc. The Materials Project website (https://next-gen.materialsproject.org/) reports that the material is synthesizable but not stable.  Whether stable or not, this material is of interest for our study. Layered NiO$_2$ has a similar structure to birnessite, where Ni ions are in a +4 OS with completely occupied $\mathrm{t_{2g}}$ states, and empty $\mathrm{e_g}$ states \cite{Carter-2019}. 

\begin{figure*}
    \centering
     \resizebox{1.0\textwidth}{!}{\includegraphics{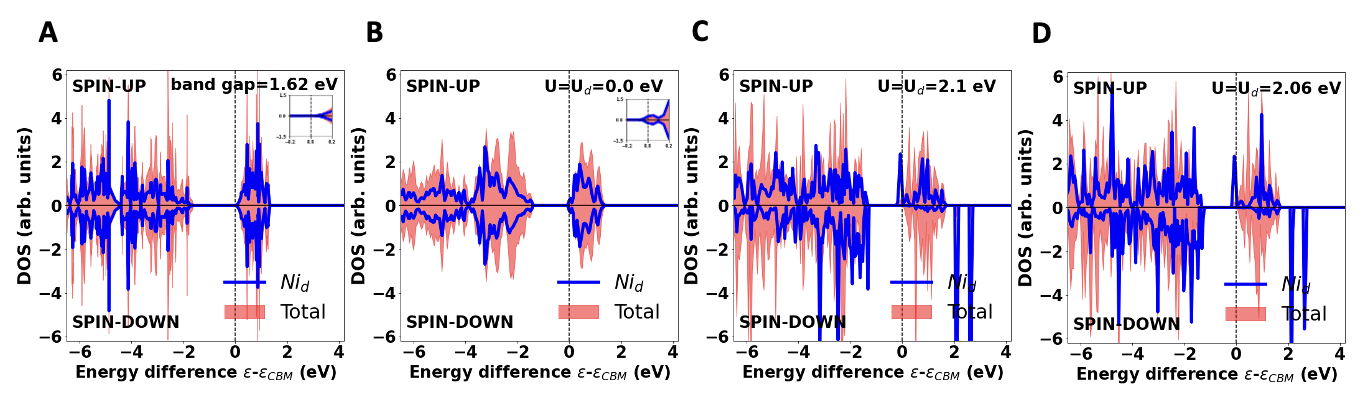}}
    \caption{r$^2$SCAN+rVV10+U+$\mathrm{U_d}$  spin-resolved density of states per atom in (A) pristine NiO$_2$ with Ni(IV) ions with U=$\mathrm{U_d}$=2.1 eV and in a single K-intercalated NiO$_2$ with (B) U=$\mathrm{U_d}$ =0.0 eV, (C) U=$\mathrm{U_d}$=2.1 eV, (D) U= $\mathrm{U_d}$ =2.06 eV. The geometry was optimized in r$^2$SCAN+rVV10+U+$\mathrm{U_d}$, which makes the interlayer spacing $\approx$ 5.42 Å. Here we can see one localized occupied state and several localized unoccupied states on the defect Ni(III) cation. The insets in Figs. A and B show the DOS near the 0 eV region. The inset in Fig. B shows that for U=$\mathrm{U_d}$=0 the extra electron from the intercalated K goes to the bottom of the conduction band, making a semi-metal.}
    \label{fig:figure5}
\end{figure*}

We began the DOS calculation using the  r$^2$SCAN+rVV10+U+$\mathrm{U_d}$ functional for both pristine and K-intercalated NiO$_2$ systems. The r$^2$SCAN U value  for Ni ions was recently determined by Swathilakshmi et al. \cite{Swathilakshmi-2023}. For the pristine NiO$_2$, r$^2$SCAN+rVV10+U+$\mathrm{U_d}$  with U=$\mathrm{U_d}$=2.1 eV, correctly predicts a non-magnetic ground state with a band gap of 1.62 eV, as shown in Fig. \ref{fig:figure5}A. As in the birnessite case, we expect that adding an extra K atom between layers would transfer an electron from the inserted K atom to a Ni site, forming a defect. The defected Ni site would undergo Jahn-Teller distortion, localizing the electron in the e$_g$ state, as shown in Fig. \ref{fig:figure7}B.

\begin{figure*}
    \centering
    \resizebox{1.0\textwidth}{!}{\includegraphics{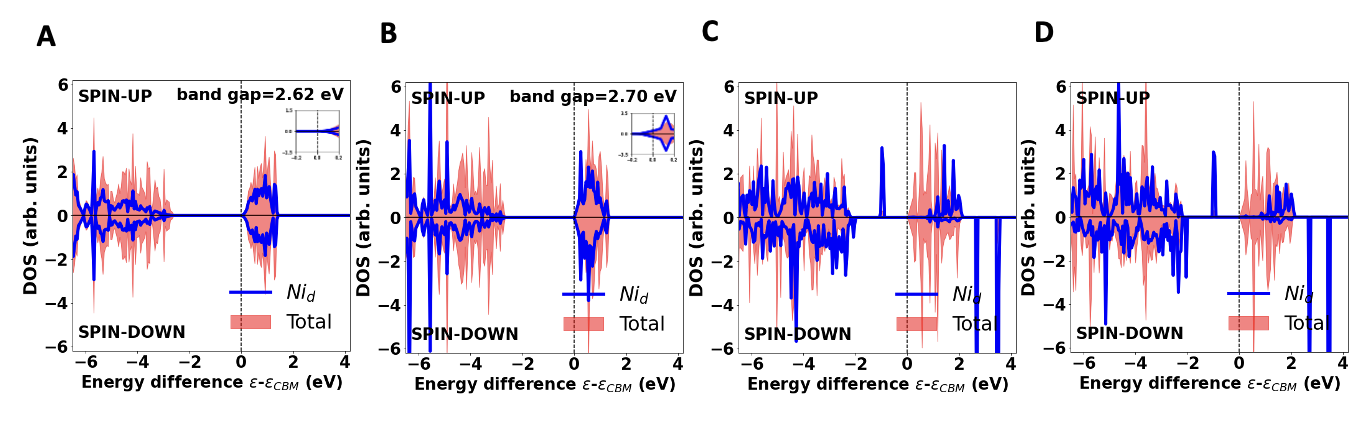}}
    \caption{HSE06+D3  spin-resolved density of states per atom using r$^2$SCAN+rVV10+U+$\mathrm{U_d}$ geometry of Fig. \ref{fig:figure5} in  (A) pristine NiO$_2$ with U=$\mathrm{U_d}$=2.1 eV and a single K-intercalated NiO$_2$ with (B) U=$\mathrm{U_d}$=0.0 eV  and (C) U=$\mathrm{U_d}$=2.1 eV,  (D) U= $\mathrm{U_d}$=2.06 eV. The insets in Figs. A and B show the DOS near the 0 eV region. The inset in Fig. B shows that for U=$\mathrm{U_d}$=0 the extra electron from the intercalated K goes to the bottom of the conduction band, making a semi-metal.}
    \label{fig:figure6}
\end{figure*}
\begin{figure*}
    \centering
    \resizebox{0.8\textwidth}{!}{\includegraphics{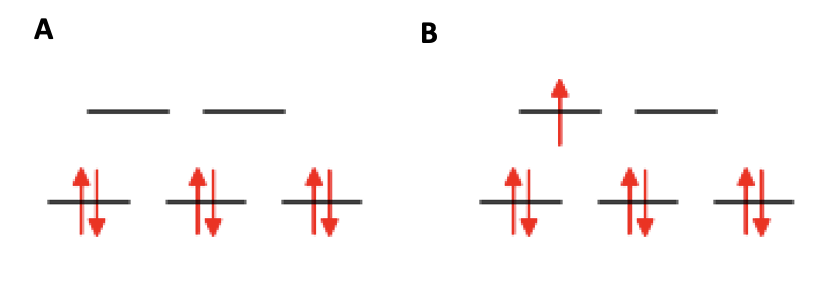}}
    \caption{d-orbital splitting (A) in Ni(IV) in pristine NiO$_2$ demonstrating filled t$_{2g}$ states and empty e$_g$ states and (B) in Ni(III) in a single K-intercalated NiO$_2$}
    \label{fig:figure7}
\end{figure*}
For a K-intercalated NiO$_2$, we initially used U=$\mathrm{U_d}$=0.0 eV to observe how r$^2$SCAN+rVV10 without U correction performs for this system, and we obtained the DOS as plotted in Fig. \ref{fig:figure5}B. Again, this choice of U and $\mathrm{U_d}$ failed to describe the defect state, as can be seen from the inset in Fig. \ref{fig:figure5}B. The extra electron goes to a delocalized state at the bottom of the conduction band, with fractional occupation on each Ni ion in the supercell, and no individual geometric distortion is observed at any nickel atom. We then used the available U value by setting U=$\mathrm{U_d}$=2.1 eV and obtained the DOS, as plotted in Fig. \ref{fig:figure5}C. This choice of U values forms a defect at one of the Ni sites, evidenced by the expected geometric distortion (see Fig. S3 B). The polaron is localized and appears in the gap just below the CB minimum. The formation of the polaron at the VB edge indicates that a K -intercalated NiO$_2$ could show OER catalytic activity similar to that of birnessite by lowering the overpotential. Most of the charge from the K atom is transferred to this defect site, which has a magnetic moment of 0.786 $\mathrm{\mu_B}$ . However, we observe a Ni site in another layer picking up a small but nonzero magnetic moment of 0.137 $\mathrm{\mu_B}$, with the magnetic moment of all remaining Ni sites smaller than 0.03 $\mathrm{\mu_B}$. This suggests that we might need different values of U and $\mathrm{U_d}$ for r$^2$SCAN+rVV10+U+$\mathrm{U_d}$ to accurately describe charge transfer to the  defect site in a K-intercalated NiO$_2$ system. 

The determination of U=2.1 eV for Ni for r$^2$SCAN involves Ni in +2 and +3 states \cite{Swathilakshmi-2023}, with an average OS of +2.5. Since the precise U values for Ni(III) and Ni(IV) states are not available, we set $\mathrm{U_d}$=2.1 eV for NI(III) as the Ni (III) OS is close to +2.5 and searched for different U values for Ni(IV). We observed that increasing the U value beyond 2.3 eV transfers more and more charge to a Ni site in a layer different from that of the defect site. So, we lowered the U values (U=1.0 eV and 0.0 eV ) and found that lowering the U values slightly improves the solution by lowering the magnetic moment of a Ni site in another layer to 0.06 $\mathrm{\mu_B}$ without significant change in magnetic moment of the defect site. The DOS plot for U=0.0 eV and $\mathrm{U_d}$=2.1 eV do not look different from Fig. \ref{fig:figure5}C. We also increased the $\mathrm{U_d}$ value, keeping U fixed at 2.1 eV, and observed an increase in the magnetic moment of the defect site, reaching 1 $\mathrm{\mu_B}$ for U=3.8 eV. These are the magnetic moments due to the spin density inside the Wigner-Seitz sphere. It is important to note that the magnetic moment of the defect site 1.0 does not guarantee the full transfer of an electron from the intercalated K atom to the defect site, as this is a magnetic moment inside the Wigner-Seitz sphere, which does not reflect the actual magnetic moment of the defect site. However, this analysis shows that one can adjust the U and $\mathrm{U_d}$ values for more charge transfer to the defect site where necessary.

We also determined the U value of 2.06 eV for the Ni(III) ion by equating the HSE06+D3 magnetic moment to the r$^2$SCAN+rVV10+U magnetic moment, as discussed in the SM \cite{SM-defect}. As this value is not very different from 2.1 eV, we got a similar r$^2$SCAN+rVV10+U+$\mathrm{U_d}$ result with U=$\mathrm{U_d}$=2.06 eV compared to U=$\mathrm{U_d}$=2.1 eV. Fig. \ref{fig:figure5}D shows r$^2$SCAN+rVV10+U+$\mathrm{U_d}$ DOS with U=$\mathrm{U_d}$=2.06 eV.

Similar to birnessite, the d-states corresponding to Ni(III) in the VB are shifted to higher energy, as shown in Figs. \ref{fig:figure5}C and \ref{fig:figure5}D, and there is spin symmetry breaking on the total DOS in  the CB. Here we do not claim U=0.0 eV to be the precise U value for Ni(IV) ions, but we suspect that U could be different than $\mathrm{U_d}$ for a K-intercalated NiO$_2$ system.

We also did HSE06+D3 calculations using the r$^2$SCAN+rVV10+U+$\mathrm{U_d}$ geometry, and the results are shown in Fig. \ref{fig:figure6}. We observe an increase in the band gap of pristine NiO$_2$, as shown in Fig. \ref{fig:figure6}A. For a single K-intercalated NiO$_2$, due to the absence of a distorted nickel center, we do not observe polaron formation for U=$\mathrm{U_d}$=0.0 eV, similar to r$^2$SCAN+rVV10+U+$\mathrm{U_d}$ as shown in Fig. \ref{fig:figure6}B; however, we observed an increase in the band gap. We also performed HSE06+D3 calculations for U=$\mathrm{U_d}$=2.1 eV  and for U=$\mathrm{U_d}$=2.06 eV and obtained DOS as shown in Figs. \ref{fig:figure6}C, and \ref{fig:figure6}D, respectively. We observe an increase in the band gap and a shift of the polaron peak deep into the band gap region compared to the corresponding r$^2$SCAN+rVV10+U+$\mathrm{U_d}$ DOS. We also observe other effects like shifting of Ni(III) d-states to higher energy and spin symmetry breaking on the total DOS in the CB. 

Both methods, full r$^2$SCAN+rVV10+U+$\mathrm{U_d}$ and HSE06+D3 using the r$^2$SCAN+rVV10+U+$\mathrm{U_d}$ geometry with proper U and $\mathrm{U_d}$ values, successfully describe defects in a single  K-intercalated NiO$_2$. Among the methods employed to study defect in a K-intercalated NiO$_2$, full r$^2$SCAN+rVV10+U+$\mathrm{U_d}$ is notably superfast.

\subsection{Layered $\mathrm{KCoO_2}$}
We studied one final related layered material: KCoO$_2$, which has a hexagonal crystal structure with space group P6$_3$/mmc, similar to the above materials MnO$_2$ and NiO$_2$. The Materials Project website (https://next-gen.materialsproject.org/) shows a similar material, LiCoO$_2$, with space group P6$_3$mc, to be synthesizable but not stable. The electronic configuration of cobalt d-states is distinct from that of Mn and Ni, as illustrated in Fig. 1 in Ref. \cite{Ding-2021}. The optimal U value determined for Co ions for r$^2$SCAN is 1.8 eV \cite{Swathilakshmi-2023}. 

\begin{figure*}
    \centering
    \resizebox{1.0\textwidth}{!}{\includegraphics{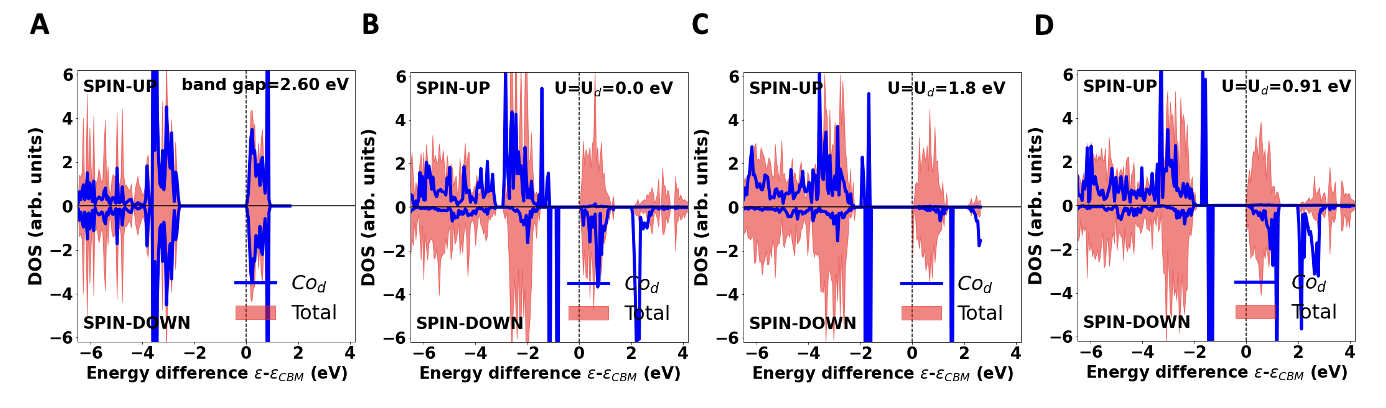}}
    \caption{r$^2$SCAN+rVV10+U spin-resolved density of states per atom  in  (A) pristine KCoO$_2$ with Co(III) ions with U=$\mathrm{U_d}$=1.8 eV, a single K-intercalated KCoO$_2$ with (B) U=$\mathrm{U_d}$=0.0 eV, (C) U=$\mathrm{U_d}$=1.8 eV, (D) U=$\mathrm{U_d}$=0.9 eV. The more complicated pattern of in-gap states could reflect a Co(II) configuration different from a single unpaired electron in an e$_g$ state. Here we see the appearance of several localized occupied states on the defect Co(II) cation. The geometry was optimized in r$^2$SCAN+rVV10+U+$\mathrm{U_d}$, which makes the interlayer spacing $\approx$ 5.98 Å. In the fundamental gap of K-intercalated $\mathrm{KCoO_2}$, we see two localized  $\mathrm{e_g}$ $\uparrow$ electrons and two localized $\mathrm{t_{2g}}$ $\downarrow$ electrons (Fig. \ref{fig:figure10}).}
    \label{fig:figure8}
\end{figure*}

First, we calculated the DOS of KCoO$_2$ using the optimal U value of 1.8 eV, as shown in Fig. \ref{fig:figure8}A. We observed symmetry in the spin resolved total DOS and Co d-states projected onto a Co site.  This suggests that Co(III) ions in KCoO$_2$ have zero spin. The zero spin of Co(III) is also reported in the work of Chen et al. \cite{Selloni-2011}. The d-state configuration of Co(III) is as shown in Fig. \ref{fig:figure10}A.

The pristine CoO$_2$ has cobalt d states in a d$^5$ configuration, and KCoO$_2$ has cobalt in a d$^6$ configuration. A K-intercalated $\mathrm{K_{1.03}CoO_2}$ would have a defect cobalt site in a d$^7$ configuration, with the remaining Co ions in a d$^6$ configuration. It is interesting to note that the Co(III) ion has no unpaired electron, but Co(II)  has three unpaired electrons due to a low-spin to high-spin crossover \cite{Sun2022,Selloni-2011, Bunting-2018,Kim-2023}. So, unlike one electron in $\mathrm{e_g}$, the state of defected Ni(III) and Mn(III), defected Co(II) has two electrons in the $\mathrm{e_g}$ states, with d-state configuration $\mathrm{t_{2g}^5e_g^2}$ (Figure \ref{fig:figure10} B, C) \cite{Sun2022}. 
\begin{figure*}
    \centering
    \resizebox{1.0\textwidth}{!}{\includegraphics{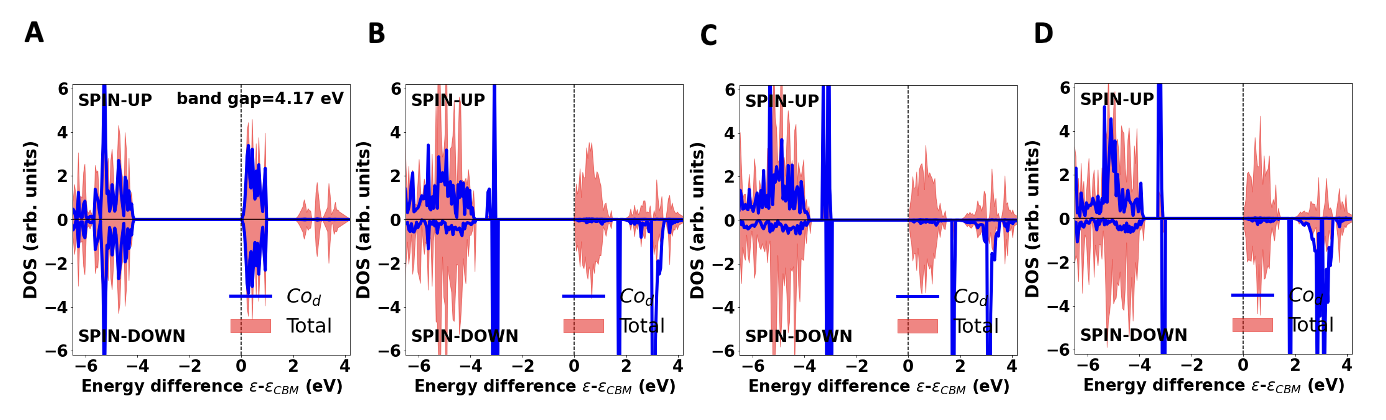}}
    \caption{HSE06+D3  spin-resolved density of states per atom using r$^2$SCAN+rVV10+U+$\mathrm{U_d}$ geometry of Fig. \ref{fig:figure8} in (A) pristine KCoO$_2$ with U=$\mathrm{U_d}$=1.8 eV, and in a single  K-intercalated KCoO$_2$ with (B) U=$\mathrm{U_d}$=0.0 eV, (C) U=$\mathrm{U_d}$=1.8 eV, and (D) U=$\mathrm{U_d}$=0.9 eV.}
    \label{fig:figure9}
\end{figure*}
\begin{figure*}
    \centering
    \resizebox{1.0\textwidth}{!}{\includegraphics{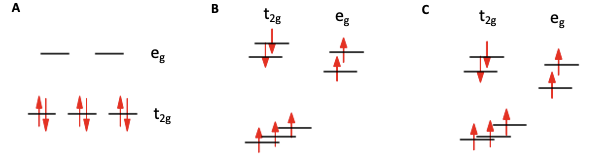}}
    \caption{Schematic diagram of energy of d-orbitals in (A) Co(III) in pristine KCoO$_2$ based on HSE06+D3 calculation using r$^2$SCAN +rVV10+U+$\mathrm{U_d}$ geometry with U=$\mathrm{U_d}$=1.8 eV, (B) Co(II)  in a single K-intercalated KCoO$_2$  based on orbital energy from r$^2$SCAN+rVV10+U+$\mathrm{U_d}$ calculation with U=$\mathrm{U_d}$=1.8 eV and (C)  Co(II)  in a single K-intercalated KCoO$_2$ based on orbital energies from the HSE06+D3 calculation using the r$^2$SCAN+rVV10+U+$\mathrm{U_d}$ geometry with U=$\mathrm{U_d}$=1.8 eV.}
    \label{fig:figure10}
\end{figure*}

We performed r$^2$SCAN+rVV10+U+$\mathrm{U_d}$ calculations to study the defects in a K- intercalated KCoO$_2$. First, we tried U=$\mathrm{U_d}$=0.0 eV to see how r$^2$SCAN+rVV10 performs without U correction, and obtained a DOS as shown in Fig. \ref{fig:figure8}B. The blue plot, which is the projected density of states (PDOS) of d-states of the Co(II) ion, shows a significant change in the d-state of Co(II) compared to the d-state of Co(III) ion.  Our calculation shows the magnetic moment of the Co(II) ion around 3 , indicating three unpaired electrons in d-states.
The presence of three unpaired electrons was previously seen in Refs. \cite{Sun2022,Selloni-2011,Kim-2023,Bunting-2018}. We observed Co(II) d-state splitting, as shown in Fig. \ref{fig:figure10}B. This explains the appearance of Co(II) d-states just above the VB maximum in both channels, absence of Co(II) d-states in the CB in the up channel and presence in the down channel. There is a breaking of spin symmetry in the DOS, due to the appearance of a spin moment on the defected Co ion. The phenomena like shifting of d-states to higher energy and spin symmetry breaking in CB are similar to those observed and explained before in Peng et al.’s study \cite{peng-2017} of birnessite. 

Then we used U=$\mathrm{U_d}$=1.8 eV in r$^2$SCAN+rVV10+U+$\mathrm{U_d}$ and obtained a DOS as shown in Fig. \ref{fig:figure8}C. We do not see much difference between the r$^2$SCAN+rVV10+U+$\mathrm{U_d}$ DOS for U=$\mathrm{U_d}$=0.0 eV and U=$\mathrm{U_d}$=1.8 eV. This indicates that the occupied Kohn-Sham orbitals are well localized and there is no fractional occupancy in r$^2$SCAN+rVV10+U+$\mathrm{U_d}$ calculations with U=$\mathrm{U_d}$=0.0 eV. However, the larger U appears to bring peaks of the defected Co(II) d-states closer together in the band gap region just above the VB maximum. 

As discussed in the SM \cite{SM-defect}, we found the U value for Co(II) to be 0.91 eV. We used U=$\mathrm{U_d}$=0.91 eV and obtained a DOS as shown in Fig. \ref{fig:figure8}D, which looks similar to Figs. \ref{fig:figure8}B and \ref{fig:figure8}C.

We tried to compare a full HSE06+D3 DOS with r$^2$SCAN+rVV10+U+$\mathrm{U_d}$  for pristine KCoO$_2$ and a K-intercalated KCoO$_2$. We obtained the KCoO$_2$ supercell by inserting a K layer for each CoO$_2$ layer. Compared to the pristine CoO$_2$ supercell, which contains 96 ions, the KCoO$_2$ supercell has 32 extra K ions, and the  K-intercalated KCoO$_2$ supercell has 33 extra K ions. Therefore, one would expect a different lattice parameter for KCoO$_2$ and a K-intercalated KCoO$_2$ , which requires the full relaxation of the supercell. We attempted to optimize the structure using the HSE06+D3 method, but it is very challenging for these systems in terms of time and resource. After a month, we chose not to proceed further. This challenge highlights one of the important reasons that motivated us to explore the r$^2$SCAN+rVV10+U+$\mathrm{U_d}$ method, highlighting the significance  of the r$^2$SCAN meta-GGA for defect studies. 

However, we calculated the HSE06+D3 DOS using r$^2$SCAN+rVV10+U+$\mathrm{U_d}$ geometry. As shown in Fig. \ref{fig:figure9}A, HSE06+D3 predicts KCoO$_2$ to be non- magnetic with an increased band gap of 4.17 eV. For a K-intercalated system with r$^2$SCAN+rVV10+U+$\mathrm{U_d}$ geometry, the HSE06+D3 DOS is shown in Figs. \ref{fig:figure9} B, C, and D for U=$\mathrm{U_d}$=0.0 eV, 1.8 eV, and 0.91 eV, respectively. Here we see that HSE06+D3 not only opens the gap between VB and CB, but it also brings the peaks of the defected Co(II) d-states closer together in the band gap region just above the VB maximum. We observed Co(II) d-states splitting as shown in Fig. \ref{fig:figure10}C for the HSE06+D3 calculation using the r$^2$SCAN+rVV10+U+$\mathrm{U_d}$ geometry with U=$\mathrm{U_d}$=1.8 eV.

Fig. \ref{fig:figure10} shows our electronic configurations for Co(III) and Co(II) in octahedral or nearly octahedral coordination with oxygen. For Co(II), we have seven d electrons. There are three $\mathrm{t_{2g}}$ orbitals of each spin, and two $\mathrm{e_g}$ orbitals of each spin, and we can put no more than one electron into each spin orbital. Our output tells us, from the relative energy order of each occupied d state of each spin, that the net spin is nearly that of three spin-up electrons, and that the lowest-energy orbitals are spin-up while the highest-occupied orbital is spin-down,  but not whether a 
given spin orbital is $\mathrm{t_{2g}}$ or $\mathrm{e_g}$. So we begin by putting three spin-up electrons into $\mathrm{t_{2g}}$ orbitals, to minimize the electronic repulsion. We assume that the $\mathrm{e_g}$ electrons prefer to be spin-up, to take advantage of an exchange interaction with the three up-spin $\mathrm{t_{2g}}$ orbitals, and that spin-up $\mathrm{e_g}$ levels will compete in energy with the spin-down $\mathrm{t_{2g}}$ levels. Then the configuration will be the one shown in Fig.
\ref{fig:figure10} B and C, $\mathrm{(t_{2g}\uparrow)^3 (t_{2g}\downarrow)^2 (e_g\uparrow)^2}$. This result agrees with the $\mathrm{(t_{2g})^5 (e_g)^2}$ configuration found for Co(II) in $\mathrm{Li_2CoO_2}$ in Ref. \cite{Sun2022}. This electronic configuration is also supported by the total charge density plots (see Fig. S5) of the four localized orbitals that appear in the band gap region of $\mathrm{K_{1.03}CoO_2}$.

Regardless of coordination, Co(II) has been found 
to have a net spin nearly equal to that of three spin-up
electrons in Refs. \cite{Selloni-2011,Sun2022,Bunting-2018,Kim-2023}, but the distribution over
$\mathrm{t_{2g}}$ and $\mathrm{e_g}$ of course depends upon coordination and other details of the nearby environment.

Fig. S4 B of the SM \cite{SM-defect} shows the Jahn-Teller distorted tetrahedron around a defect Co(II) ion, with two of the Co(II)-O bonds about 10\% longer than the remaining four. The angles are also distorted.

\section{Conclusions}
Defects in solids can be utilized to tune the electronic structure of materials, impacting applications in various fields. Defects can be investigated using first-principles calculations with DFT. However, DFAs introduce SIE that could vary depending on the element, system, site, and oxidation state, hindering the performance of DFAs. Hybrid functionals mitigate SIE by mixing exact exchange , but at the expense of computational cost, successfully describing defects in many materials. The computational cost of hybrid functionals increases significantly for a SCF (self-consistent field) calculation compared to meta-GGAs like r$^2$SCAN. Defect studies require large supercells, making SCF calculations very expensive. Ionic relaxation calculations for defect studies involve many SCF cycles, making the process cumbersome for a hybrid functional to optimize the structure. 

Here we demonstrate that the r$^2$SCAN+rVV10+U+$\mathrm{U_d}$ method can successfully predict the existence of localized polaronic states near defects, with literature values of U and $\mathrm{U_d}$. The method is much faster than hybrid functionals but but still predicts localized polaronic states in band gaps. Using the r$^2$SCAN+rVV10+U+$\mathrm{U_d}$ geometry enables the completion of hybrid functional (or even G0W0) calculations in a reasonable time and with fewer resources. This approach makes hybrid functional calculations feasible for larger systems with transition-metal ions. 

For fixed geometries, we have found the same number of occupied states localized on the defect transition-metal ion in the energy gap below the bottom of the conduction band, whether we use HSE06 ($\mathrm{\alpha = 0.22 \ or \ 0.25}$) or r$^2$SCAN+rVV10+U+$\mathrm{U_d}$ (in a range of U and $\mathrm{U_d}$ around the literature values \cite{Swathilakshmi-2023}), for layered $\mathrm{MnO_2}$ (Figs. 3-4), $\mathrm{NiO_2}$ (Figs. 6-7), or $\mathrm{KCoO_2}$ (Figs. 9-10). We have found the same whether we use
the HSE06 or the r$^2$SCAN+rVV10+U+$\mathrm{U_d}$
geometries, for $\mathrm{MnO_2}$ (Figs. 2 and 4). (When HSE06 is used without a dispersion correction, the interlayer spacing must be constrained to a reasonable value.) Fig. 11 shows the same density-of-states orbital ladder from both functionals in $\mathrm{KCoO_2}$. This level of agreement is found, even though the fundamental band gaps are significantly smaller for r$^2$SCAN+rVV10+U+$\mathrm{U_d}$ than for HSE06. In the SM \cite{SM-defect}, we also find good agreement with HSE06+D3 for the magnetic moments of the transition-metal ions and for the lattice constants of the pristine materials. Jahn-Teller distortions of the metal-oxygen bond lengths at the defect are also shown in the 
SM \cite{SM-defect}. The feasibility of other quantitative comparisons, including comparisons with experiment, will be considered in our future work.

This method can expedite the study of defects in materials. The r$^2$SCAN+rVV10+U+$\mathrm{U_d}$ method can also be employed in systems without defects where ions of the same species are in different OS.

We have studied the layered oxides MnO$_2$, NiO$_2$, and KCoO$_2$, both in the pristine state and with one additional K atom intercalated between layers in a supercell. Inexpensive r$^2$SCAN+rVV10+U+$\mathrm{U_d}$ equilibrium geometries have been used for electronic structure calculations with r$^2$SCAN+rVV10+U+$\mathrm{U_d}$ and with the expensive HSE06 hybrid functional. For K-intercalated MnO$_2$ and NiO$_2$, we find no localized e$_g$ state on the defected transition metal ion for U=$\mathrm{U_d}$=0, but we find one such state for standard positive U values. This state, in the gap between conduction and valence bands, accepts the electron donated by the intercalated K. (For U=$\mathrm{U_d}$=0, the extra electron from the intercalated K goes into the bottom of the conduction band, making the intercalated material a semi-metal. The extra electron is then delocalized over the supercell, with fractional occupation on each transition-metal ion.)

For K-intercalated KCoO$_2$, both for U=$\mathrm{U_d}$=0 (standard semilocal r$^2$SCAN without a nonlocal +U self-interaction correction) and for standard positive U values, we seem to find two majority-spin and two minority-spin occupied localized e$_g$ states in the gap between valence and conduction bands, and three empty localized minority-spin t$_{2g}$ states in the gap above the conduction band. This surprising result is consistent with a dramatic change in the electronic configuration (Fig. \ref{fig:figure10}) from the undefected Co(III) ions to the defected Co(II) ion, as reported previously in Ref. \cite{Selloni-2011}.

The r$^2$SCAN+rVV10+U+$\mathrm{U_d}$ approach is successful, but it has limitations. It requires material-dependent parameters U, and those are typically available for only a few oxidation states of each transition metal. For cases where a literature value of U is unavailable, we have suggested (in the
SM \cite{SM-defect}) a way to find a value by matching the HSE06 magnetic moment of the transition-metal ion (when it is non-zero). The HSE06 calculation could be made for a pristine material, avoiding large supercells. An interesting approach to U estimation, based on an ensemble of low-and high-spin states for a given transition metal ion and oxidation state, was suggested by Ref. \cite{Albavera-Mata}.

\section{Computational Details}
First-principles calculations were performed with the projector-augmented wave method \cite{paw-1994}, implemented in the VASP code \cite{Kresse-1994,Kresse-1999}, version 5.4.4.. A 4 x 4 x 1 supercell was used to simulate defects in layered TMOs. For all supercell calculations, a 2 x 2 x 2 $\Gamma$-centered Monkhorst-Pack k mesh \cite{Pack-1976} was used. 

For r$^2$SCAN+rVV10+U+$\mathrm{U_d}$ calculations, a 500 eV cutoff for the plane waves was used. In all r$^2$SCAN+rVV10+U+$\mathrm{U_d}$ calculations, the cell volume was relaxed with the ISIF=3 setting until forces converged to less than 0.03 eV/Å and energy converged to less than 10$^{-6}$ eV. To conduct r$^2$SCAN+rVV10+U+$\mathrm{U_d}$ calculations, we employed the simplified rotationally invariant framework developed by Dudarev et al. \cite{Dudarev-1998}.

For Fig. \ref{fig:figure1}, we used an exact mixing parameter $\alpha$ of 0.22 in the hybrid functional HSE06 \cite{peng-2017,HSE06-Krukau-2006}, as determined from the Generalized Koopman’s Condition (GKC) method \cite{Lany-2009,Lany-2011,Lany-2012,Lany-2015}. For all the remaining HSE06+D3 calculations, we used the standard exact exchange mixing parameter of 0.25. A plane-wave basis with an energy cutoff of 400 eV was employed. \\

\section{Acknowledgments}
This work was supported by the U.S. Department of Energy, Office of Science Basic Energy Sciences under Award No. DE-SC0023356. We thank NERSC for computational resources. \\

\maketitle
\bibliography{references}

\end{document}